\documentclass[twocolumn]{aastex701}
\usepackage{amsmath} 
\usepackage[T1]{fontenc}

\begin{document}
\linenumbers
\title{Magnetic Pressure Dominance Stabilizes AGN Disks Against Gravitational Instability}
\author[]{Hannalore J. Gerling-Dunsmore}
\affiliation{JILA, University of Colorado and National Institute of Standards and Technology, 440 UCB, Boulder, CO 80309-0440, USA}
\affiliation{Department of Astrophysical and Planetary Sciences, 391 UCB, University of Colorado, Boulder, CO 80309-0391, USA}
\email[show]{gerlingd@colorado.edu}  

\author[0000-0003-0936-8488]{Mitchell C. Begelman}
\affiliation{JILA, University of Colorado and National Institute of Standards and Technology, 440 UCB, Boulder, CO 80309-0440, USA}
\affiliation{Department of Astrophysical and Planetary Sciences, 391 UCB, University of Colorado, Boulder, CO 80309-0391, USA}
\email[]{-}

\author[0000-0002-3771-8054]{Jacob B. Simon}
\affiliation{Department of Physics and Astronomy, Iowa State University, Ames, IA, 50010, USA}
\email[]{-}

\author[0000-0001-5032-1396]{Philip J. Armitage}
\affiliation{Center for Computational Astrophysics, Flatiron Institute, 162 Fifth Avenue, New York, NY 10010, USA}
\affiliation{Department of Physics and Astronomy, Stony Brook University, Stony Brook, NY 11794, USA}
\email[]{-}

\begin{abstract}

Magnetic effects have long been considered a possible factor in stabilizing the outer regions of active galactic nuclei (AGN) accretion disks against gravitational instability (GI). However, the computational demands of testing this hypothesis have prevented comprehensive study of this problem. Here, we present results from a suite of 6 isothermal magnetohydrodynamics (MHD) shearing box simulations, 3 initialized with strong magnetization ($\beta^{\rm{mid}}_{0} = p_{\rm{gas}} / p_{\rm{mag}} = 10^{2.5}$) and 3 initialized with weak magnetization ($\beta^{\rm mid}_{0} = 10^{4}$). For each magnetization, we performed simulations with both strong ($Q_{0} = 1.0$) and weak ($Q_{0} = 10.0$) self-gravity, where $Q_{0} = \frac{c_{\rm{s}}\Omega}{\pi G \Sigma_{0}}$ is the Toomre stability parameter; we also performed pure MHD simulations for comparison. We find that our strongly magnetized disk stabilized against GI after initialization to critical stability against GI, while our corresponding weakly magnetized disk did not. We show that the strongly magnetized, strongly self-gravitating disk became dominated by magnetic pressure, which led to its stabilization. 
\end{abstract}




\section{Introduction}\label{sec:intro}
\indent The fueling of active galactic nuclei (AGN) via accretion disks \citep{LyndenBell1969} presents multiple open theoretical problems. A successful AGN accretion physics model must both allow the transport of matter through the accretion disk sufficiently quickly to power AGN at the highest observed redshifts, and explain how (or whether) the accretion disk can exist beyond the broad line region (BLR). Standard disk models  predict that outside the BLR, the accretion disk should become gravitationally unstable via the Toomre instability, and fragment, ultimately collapsing into stars \citep{ShlosmanBegelman1987,ShlosmanBegelman1989,Goodman2003}. Star formation within AGN disks is a plausible physical process that provides an efficient channel for the formation of stars and compact objects close to the Galactic Center \citep{Levin2007} and other supermassive black holes \citep{McKernan2018,Dittmann2020}. For most AGN, however, there is a lack of observational evidence for star formation immediately outside the BLR, and obvious tension between the requirement that the disk must be able to transport matter from large radii to power the observed AGN signatures and the prediction that it typically fragments into stars \citep{ShlosmanBegelman1987,ShlosmanBegelman1989,Goodman2003}. Thus, the question of how disks survive into the self-gravitating region is central to understanding diverse problems including the origin of the Galactic Center stellar population, the co-evolution of supermassive black holes with their galaxies, and the relationship between AGN and starburst galaxies \citep{Thompson2005}.

\indent Since the introduction of the $\alpha$-disk model of accretion disks \citep{ShakuraSunyaev1973}, the central problem of accretion theory has been to isolate the processes that drive outward angular momentum transport and inward matter transport. In AGN and other well-ionized disk systems, the magnetorotational instability \citep[MRI:][]{BalbusHawley91} has long been studied as a major source of turbulence and effective viscosity. More recently, it has been suggested that the subset of MRI-unstable disk solutions that possess a moderate amount of net poloidal (vertical) magnetic flux are of particular observational interest. In a disk with sufficiently strong magnetization (with the midplane net vertical field $B_{\rm 0z}$ satisfying  $\beta^{\rm{mid}}_{0} \equiv 8\pi p_{\rm gas}/B_{\rm 0z}^2 \lesssim 10^{3}$, where $p_{\rm gas}$ is the gas pressure), simulations \citep{BaiStone2013,Salvesen2016a} show that magnetic pressure throughout the disk exceeds the gas pressure, developing magnetic domination (total plasma $\beta \lesssim 1$). 

\indent Angular momentum transport due to the self-gravity of the disk can also act as an effective viscosity, although when too strong it leads to gravitational instability (GI), which in turn drives disk fragmentation \citep{Toomre1964,Paczynski1978,KolykhalovSunyaev1980,Gammie2001}. The importance of disk self-gravity is quantified by the Toomre Q-parameter, $Q = {c_{\rm{s}}\Omega}/{(\pi G \Sigma)}$, where $c_{\rm{s}}$ is the sound speed, $\Omega$ the angular velocity and $\Sigma$  the surface density \citep{Toomre1964}. Instability occurs when $Q < Q_{\rm crit} \sim 1$. In non-self-gravitating $\alpha$-model disks, $Q \propto \alpha c_{\rm{s}}^{3}/ \dot{M}$, so self-gravity inevitably becomes important at high accretion rates and low disk sound speed, i.e., at sufficiently large distances from the black hole. These considerations underlie the standard qualitative model of AGN disk structure in which the inner region is non-self-gravitating and MRI unstable, an intermediate zone is unstable to both self-gravity (in the ``stable angular momentum transport" regime) and the MRI, and an outer part is violently unstable to self-gravitating collapse. 

\indent The MRI and GI are predicted to co-exist in the outer regions of both protoplanetary \citep{Kratter2016} and AGN disks. In early work, \cite{Fromang2004} showed that the interaction between these instabilities, in disks where they co-exist, is non-trivial --- the stress of the combined instability is not the sum of the individual instabilities. The influence of the MRI on fragmentation, in the limit of strong GI, is likely to be similarly complex. Recent work has largely focused on 
conditions more appropriate to protoplanetary disks than AGN disks \citep{RiolsLatter2018a,RiolsLatter2018b,LP22,LP23,Deng2012}. While this work is useful in suggesting general trends to expect, the differences between protoplanetary disks and AGN disks are substantial and motivate focused study of the AGN case. Most obviously, non-ideal magnetohydrodynamic (MHD) effects (Ohmic diffusion, the Hall effect, and ambipolar diffusion) are dominant at almost all radii in protoplanetary disks \citep[e.g.][]{Armitage2011}, including on the large scales where GI occurs. Non-ideal physics is substantially less important in AGN \citep{MenouQuataert2001}. 

\indent In this paper, we study the interaction of the two instabilities in conditions appropriate to AGN disks, and particularly to AGN disks at the onset of GI. We employ an isothermal equation of state (and therefore do not include cooling) to approximate the thermal conditions for a disk that is strongly irradiated and at most moderately optically thick. The key numerical challenge is the scale separation between the most unstable linear modes of the MRI (small) and of GI (large). We prioritize proper resolution of the MRI, due to its known importance in AGN disks. In order to determine the impact of magnetic pressure on disk stabilization, we select a weak magnetization ($\beta^{\rm{mid}}_{0} = 10^{4}$) which is insufficient to drive magnetic pressure dominance, and a strong magnetization ($\beta^{\rm{mid}}_{0} = 10^{2.5}$), which has been shown to be sufficient to drive magnetic pressure dominance in pure MHD studies (\citealt{BaiStone2013,Salvesen2016a}). In Section \ref{sec:methods}, we outline our simulation set-up and diagnostics. In Section \ref{sec:results}, we present the results of our simulations and discuss their physical implications. In Section \ref{sec:analysis}, we show what our results imply for disk stability. In Section \ref{sec:disussion}, we explore broader implications of this study, and we conclude in Section \ref{sec:conclude}.


\section{Method of Approach}\label{sec:methods}
\subsection{Governing Equations}\label{subsec:eqs}
\indent We adopt a shearing box configuration for computational efficiency, and employ {\sc Athena++} to solve the ideal MHD equations and Poisson equation:
\begin{equation}
    \label{eqn:mhdmasseq}
    \frac{\partial \rho}{\partial t} + \nabla \cdot (\rho \mathbf{v}) = 0
\end{equation}
\begin{equation}
    \label{eqn:mhdmomeq}
    \begin{split}
        \frac{\partial \rho \mathbf{v}}{\partial t} + \nabla \cdot (\rho \mathbf{v} \mathbf{v} - \mathbf{BB} + p_{\rm gas} + B^{2}/2) = \\
    -\rho \nabla \Phi - 2\rho\Omega \mathbf{\hat{z}} \times \mathbf{v} + 2q\rho\Omega^{2}x\mathbf{\hat{x}} - \rho\Omega^{2}z\mathbf{\hat{z}}
    \end{split}
\end{equation}
\begin{equation}
    \label{eqn:mhdinductioneq}
    \frac{1}{c}\frac{\partial \mathbf{B}}{\partial t} - \nabla \times (\mathbf{v} \times \mathbf{B}) = 0
\end{equation}
\begin{equation}
    \label{eqn:fishy}
    \nabla^{2} \Phi = 4\pi G \rho
\end{equation}
\noindent where $\rho$ is the gas density,  $\mathbf{v}$ is the 3D flow velocity, $p_{\rm{gas}} = \rho c_{\rm{s}}^2$ is the isothermal gas pressure defined by the constant isothermal sound speed $c_{\rm{s}}$, $\Phi$ is the self-gravitational potential, $\Omega$ is the orbital frequency of the disk, $q = 3/2$ is the Keplerian shear parameter, and $\mathbf{B}$ is the magnetic field. We solve our governing equations using the standard Godunov method in {\sc Athena++} and use fast Fourier transforms for solving the Poisson equation (Eq.~\ref{eqn:fishy}).

For readers unfamiliar with shearing boxes, we briefly detail the implementation of the shearing approximation. We use an orbital advection scheme to separately evolve the background shear flow and the velocity fluctuations (choosing the value $q = 3/2$ for a Keplerian disk): 
\begin{align}
    \mathbf{v_{shear}} &= -q \Omega x \hat{j}\\
    \mathbf{v} &= \mathbf{v}' - \mathbf{v_{shear}}
\end{align}
where $\mathbf{v}$ is the flow velocity of our segment of the disk, $\mathbf{v_{shear}}$ is the velocity due to the differential rotation of the disk, and $\mathbf{v}'$ is the velocity fluctuation.

We make the usual choice of periodic boundary conditions for the toroidal ($y$) boundaries and shearing-periodic boundary conditions for the radial ($x$) boundaries. We also opt for periodic boundary conditions for the vertical ($z$) boundaries. While a somewhat atypical choice for shearing box studies of AGN disks, periodic vertical boundary conditions improve the stability of our simulations. Outflow boundary conditions are the more typical (and physically accurate) choice, and necessary for investigating many accretion phenomena. However, for this study, we are primarily interested in the interaction of the MRI and GI within a scale height of the disk midplane. Due to our box size, the vertical boundaries are sufficiently far from the region of interest as to make our choice acceptably unimportant. Further, recent work \citep{Squire2025Rapid} indicates that toroidal field escape is likely minimal, and that the choice of boundary conditions has little impact on the midplane region (within one density scale height $\pm H_{\rm{\rho}}$, defined in Section \ref{subsec:defs}), where we focus most of our analysis. Due to the intense computational demands of the simulations, we use a Harten-Lax-van Leer-Einfeldt (HLLE) Riemann solver for its improved numerical stability.

\begin{deluxetable*}{llllll}
    \tablewidth{0pt}
    \tablecaption{Simulation parameters: $Q_{\rm{MRI,z}}$ is the MRI quality factor for the vertical component of the magnetic field, given by Eq.~\ref{eqn:qmri}. \label{tab:simparam}}
    \tablehead{
    \colhead{$\beta^{\rm{mid}}_{0}$} & \colhead{$Q_{0}$} & \colhead{domain} & \colhead{grid} & \colhead{$Q_{\rm{MRI,z}}$}
}
\startdata
$10^{2.5}$ & 1.0 & $6H \times 6H \times 6H$ & $128 \times  128 \times 128$ & 16.52 \\
$10^{2.5}$ & 10.0 & $6H \times 6H \times 6H$ & $128 \times 128 \times 128$ & 16.52 \\
$10^{2.5}$ & MHD & $6H \times 6H \times 6H$ & $128 \times 128 \times 128$ & 16.52 \\
$10^{4}$ & 1.0 & $6H \times 6H \times 6H$ & $256 \times 256 \times 256$ & 5.87 \\
$10^{4}$ & 10.0 & $6H \times 6H \times 6H$ & $128 \times 128 \times 128$ & 2.94 \\
$10^{4}$ & MHD & $6H \times 6H \times 6H$ & $128 \times 128 \times 128$ & 2.94 \\
\enddata
\end{deluxetable*}

\subsection{Simulation Setup}\label{subsec:simsetup}
\indent We choose units such that $c_{\rm s} = \Omega = 0.001$. We then define our scale height $H = c_{\rm{s}}/\Omega = 1.0$ and initial surface density as $\Sigma_{0} = \rho_{0} H \sqrt{2\pi}$  \citep{Carrera2015}. Our gravitational constant is then calculated using $Q = \frac{c_{\rm{s}} \Omega}{\pi G \Sigma_{0}}$. In these units, the Jeans length (and wavelength of the marginally stable mode) is $L_{\rm{J}} = \frac{c_{\rm{s}}^{2}}{G \Sigma} = \pi H$ for $Q = 1.0$. We choose a box size of $6H \times 6H \times 6H$ to allow development of multiple GI modes while maintaining computational feasibility.

We determined the resolution for our simulations using the MRI quality factor described in \cite{Hawley2011}: 
\begin{equation}
    \label{eqn:qmri}
    Q_{\rm{MRI,z}} = \frac{2\pi v_{\rm{A}}}{\Omega \Delta z}
\end{equation} 
where $v_{\rm{A}} = \frac{B}{\sqrt{\mu_{0}\rho}}$ is the Alfvén speed of the plasma and $\Delta z$ is the vertical grid zone size. All three of our strongly magnetized ($\beta^{\rm{mid}}_{0} = 10^{2.5}$) simulations, as well as our weakly self-gravitating ($Q_{0} = 10.0$) and pure MHD weakly magnetized ($\beta^{\rm{mid}}_{0} = 10^{4}$) simulations were run at a resolution of $128 \times 128 \times 128$. This resolution yields $Q_{\rm{MRI,z}} = 2.94, 16.52$ for $\beta^{\rm{mid}}_{0} = 10^{4}, 10^{2.5}$, respectively. While our strongly magnetized cases are thus fully resolved, this resolution is on the border of acceptable for our weakly magnetized disks.

The quality factor for our weakly magnetized simulations, $Q_{\rm{MRI,z}} = 2.94$, is a typical $Q_{\rm{MRI,z}}$ value for studies of this nature \citep{RiolsLatter2018a,RiolsLatter2018b} and requires significantly less computing time. However, under-resolving the MRI in our weakly magnetized, $Q_{0} = 1.0$ simulation would prevent us from determining whether the MRI in a disk that is not magnetic pressure-dominated can suppress fragmentation in the weakly magnetized case. Consequently, we increased the resolution to $256 \times 256 \times 256$ for $\beta^{\rm{mid}}_{0} = 10^{4}, Q_{0} = 1.0$, yielding $Q_{\rm{MRI,z}} = 5.87$. While not ideal, this $Q_{\rm{MRI,z}}$ value is sufficient for us to confirm that the outcome of the weakly magnetized $Q_{0} = 1.0$ case is not determined by the resolution.  This information is summarized in Table~\ref{tab:simparam}.

Non-self-gravitating, isothermal AGN disks are expected to be in hydrostatic equilibrium (HSE), with vertical stratification characterized by a Gaussian density profile $\rho(z) = \rho_{0} \exp\left(-{\frac{z^{2}}{2H^{2}}}\right)$, where we set the initial peak density $\rho_{0} = 1$ for convenience. As is standard practice, we introduce small, random perturbations to the density. In the presence of self-gravity, HSE is attained with a somewhat altered vertical density profile, as shown in \cite{Shi_2014}; however, as the deviations from a Gaussian profile are small, we opt to employ an initial Gaussian profile for all simulations to enable a direct comparison between our self-gravitating disks and our pure MHD disks. As a result of the Gaussian profile, our self-gravitating disks develop initial breathing modes, which damp out before the onset of significant density inhomogeneities. We use a density floor of $\rho_{\rm{floor}} = 10^{-4}$, as in \cite{Shi_2014}, \cite{LP22}, and \cite{LP23}. As is appropriate for the outer regions of AGN disks, we choose an isothermal equation of state. 

We initialize our simulations with a net vertical magnetic field. The degree to which AGN disks possess net vertical flux is a matter of debate. While net flux is not required for the MRI to provide effective $\alpha$-viscosity, and the means by which a net vertical field is established are still unknown, there are several reasons we opt for net vertical flux in our simulations. If the magnetization is stronger than $\beta^{\rm mid}_{0} \approx 10^3$, the disk becomes magnetically dominated \citep{BaiStone2013}. Magnetically dominated disks are robust against thermal/viscous instability \citep{BegelmanPringle2007,Sadowski2016} and are expected to be more robust against gravitational instability \citep{Pariev2003,BegelmanPringle2007,Gaburov2012}. Thus, for the purposes of determining whether the MRI can stabilize AGN disks against gravitational fragmentation, as well as determining the parameter space which permits stabilization, net vertical flux is the most promising scenario.

There are additional reasons for our choice of net vertical flux. For disks with magnetization stronger than $\beta^{\rm{mid}}_{0} \approx 10^{5}$, net vertical flux provides higher values of $\alpha$ compared to the zero net flux case \citep{Hawley1995}. Further, net vertical flux aids in wind-launching, and potentially is vital in launching sufficient AGN feedback \citep{blandfordpayne, FerreiraPelletier93a,SuzukiInutsuka2009,Fromang2013,BaiStone2013}.

\subsection{Modifications to the Toomre Q Parameter}\label{subsec:modQ}

The derivation of the classical Toomre Q-parameter includes only hydrodynamics and self-gravity. In that case, the sound speed suffices to parametrize the pressure from gas within the disk. However, in the case of a highly turbulent disk with magnetization sufficient to drive the MRI, there are additional sources of pressure that can act to counteract self-gravity. To capture these effects, we will use a modified definition of the Toomre stability parameter that includes terms to account for the pressure from the magnetic field and the turbulence driven by the MRI:   
\begin{equation}
    \label{eqn:modQ}
    Q' = \frac{\Omega_{0} \sqrt{c_{\rm{s}}^{2} + v_{\rm{A}}^2 + v_{\rm{turb}}^2}}{\pi G \Sigma_{0}}
\end{equation}
where $v_{\rm A}$ is the average Alfvén speed in the disk and $v_{\rm turb}$ is the average turbulent velocity (i.e.~the gas velocity with the shear velocity subtracted) in the disk. Such modifications are not novel to our study; their validity has been demonstrated by \cite{KimOstriker01} and \cite{Lizano_2010}. 

\subsection{Density Scale Height, Stabilization, and Fragmentation}\label{subsec:defs}

We now define several terms we will use throughout this paper: disk scale height, stabilization, and fragmentation. 

While the vertical structure of a non-self-gravitating disk is well-described by the Gaussian profile derived from hydrostatic equilibrium with a characteristic scale height $H$, in the presence of self-gravity the vertical structure changes. We define the density scale height to be $H_{\rho} = \sqrt{\int \rho z^{2} dz / \int \rho dz}$, as in \cite{Sadowski2016}. We consider a disk stabilized if within $\pm H_{\rho}$, its volume-averaged modified Toomre stability parameter, $Q'$, reaches (and sustains) a value greater than unity. 

As it would cost a significant amount of computing resources to run simulations until the disk visibly fragments, we use a more physically motivated definition of fragmentation. We consider a disk fragmented if its modified Toomre $Q'$ value plateaus at a value $Q' \leq 1.0$ (or continues to decrease) after the Maxwell stress has plateaued. In our study, the only possible source of disk stabilization is the MRI; if the Maxwell stress has plateaued, the disk has developed as much magnetic turbulence as possible. Consequently, if the turbulence is insufficient to raise the total pressure in the disk (thus stabilizing the disk against GI) when it saturates, there is no means by which the disk can stabilize. 

\subsection{Diagnostics}\label{subsec:diagnostics}

We now define the diagnostics used in our study. We define time, horizontal, and volume averages in the usual way:

\begin{equation}
    \langle X \rangle_{t} = \frac{1}{t_{\rm{f}} - t_{\rm{i}}} \int^{t_{\rm{f}}}_{t_{\rm{i}}} X dt
\end{equation}

\begin{equation}
    \langle X \rangle_{xy} = \frac{1}{L_{\rm{x}}L_{\rm{y}}} \iint X dy dx
\end{equation}

\begin{equation}
    \langle X \rangle_{V} = \frac{1}{L_{\rm{x}}L_{\rm{y}}L_{z}}\iiint_{\rm{V}} X dxdydz.
\end{equation}

Averages performed over multiple dimensions are denoted using nested brackets with corresponding subscripts, in order of averaging, e.g.~$\langle\langle X \rangle_{\rm{V}}\rangle_{\rm{t}}$ indicating averaging of the quantity $X$ first over volume, then over time. A superscript $X^{\rm{mid}}$ indicates that the quantity was evaluated at the midplane of the disk, and a subscript $X_{0}$ indicates that the quantity was evaluated at $t = 0$. For ratios such that $Z = \frac{X}{Y}$, a bar ($\Bar{Z}$) indicates the ratio of the volume averages of $X$ and $Y$, and a hat ($\hat{Z}$) indicates the ratio of the horizontal averages of $X$ and $Y$.

\section{Results}\label{sec:results}
\subsection{Overview}\label{subsec:overview}

We present results from our six simulations, with an overview of parameters shown in Table \ref{tab:simparam}. We employed two magnetization values ($\beta^{\rm{mid}}_{0} = p_{\rm{gas}}/p_{\rm{mag}} = 10^{2.5}, 10^{4}$); for both magnetization values, we performed a strongly self-gravitating simulation initialized to marginal stability against GI ($Q_{0} = 1.0$), a weakly self-gravitating simulation initialized to robust stability against GI ($Q_{0} = 10.0$), and a pure MHD simulation. We ran the pure MHD simulations and weakly magnetized, robustly stable self-gravitating simulations to 25 orbits, as pure MHD studies have found that MRI saturates by 25 orbits \citep{Salvesen2016a}. 
 
The inclusion of self-gravity significantly increased the required computational resources, especially in our strongly magnetized (and therefore highly turbulent) simulations. In order to most efficiently use computing time, simulations initialized to marginal stability ($Q_{0} = 1.0$) were terminated either upon their modified Toomre parameter exceeding the critical threshold ($Q' > 1.0$) for at least two orbits, or upon fragmentation of the disk as defined in the preceding section. The strongly magnetized, weakly self-gravitating simulation required significantly more computational resources than its weakly magnetized counterpart, due to the aforementioned computational demands of self-gravity in a turbulent system. Thus, we ran the strongly magnetized, weakly self-gravitating simulation until 10 orbits after the saturation of the toroidal magnetic field (to an end point of 15 orbits). After determining it remained converged with the strongly magnetized, pure MHD simulation, we terminated the simulation.  

Previous studies of the interaction of the MRI and self-gravity found that if a disk attains $Q \geq 1.0$ it will remain stable, even if run for hundreds of orbits \citep{LP23}. This sustained stabilization occurs despite under-resolution of the MRI, which weakens the MRI's ability to generate the magnetic pressure required to oppose the self-gravity of the disk. \cite{tsung2025doesmagneticfieldpromote} found that it is technically possible for MRI to enhance GI, but for the magnetizations employed in this study such GI enhancement is highly unlikely to occur. By this criterion, attaining and then maintaining $Q' \geq 1.0$ for an orbit adequately demonstrates that the disk has stabilized and will remain stable, and we did not compute these models further. 

In the remainder of this section, we will demonstrate that the MRI developed normally in all three of our strongly magnetized simulations, as well as in our weakly magnetized, weakly self-gravitating and pure MHD simulations.  Our weakly magnetized, strongly self-gravitating simulation ultimately fragmented, and we will show how that altered standard MRI diagnostics for that simulation. We will then show that our strongly magnetized disks became dominated by magnetic pressure, while the weakly magnetized disks did not, and discuss how magnetic pressure dominance altered the disk structure.  

\begin{figure*}
\centering
\includegraphics[width=18cm]{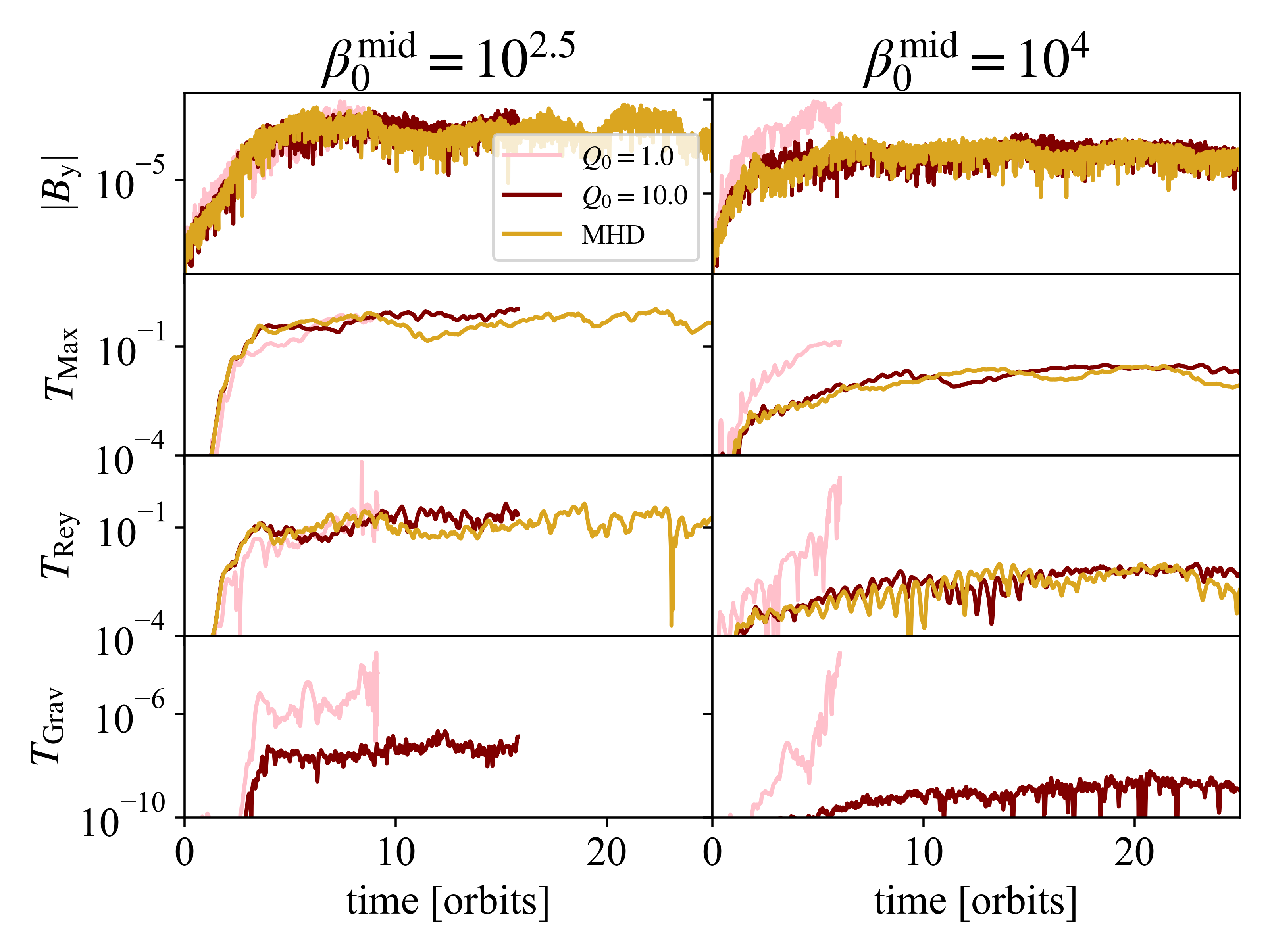}
\caption{The rows show, from top to bottom, the evolution of the volume-integrated toroidal magnetic field (absolute value taken after volume integration), Maxwell stress, Reynolds stress, and gravitational stress. In the strongly magnetized disk ($\beta^{\rm{mid}}_{0} = 10^{2.5}$) that survives initialization to $Q_{0} = 1.0$, the evolution of the Maxwell stress is similar in the pure MHD, weakly self-gravitating, and strongly self-gravitating simulations. As is expected, the evolution of the Maxwell stress tracks the development of the toroidal field. However, the weakly magnetized ($\beta^{\rm{mid}}_{0} = 10^{4}$), strongly self-gravitating simulation displays both significantly enhanced Maxwell and Reynolds stress, indicating strong magnetic and fluid turbulence prior to fragmentation of the disk. Additionally, this simulation shows an enhancement of the toroidal field (prior to fragmentation). Interestingly, the strongly magnetized, strongly self-gravitating disk does not display any significant difference from its pure MHD and weakly self-gravitating counterparts. In all cases, the gravitational stress is several orders of magnitude weaker than the Maxwell or Reynolds stress.}
\label{fig:stresses}
\end{figure*}

\subsection{MRI Diagnostic: Evolution of Stresses}\label{subsec:stresses}
\indent As is standard in studies of the MRI in accretion disks, we utilize the $r\phi$-components of the Maxwell and Reynolds stresses as diagnostics: 
\begin{equation}
    \label{eqn:maxstress}
    T_{\rm{xy,Max}} = -B_{\rm{x}}B_{\rm{y}}
\end{equation}
\begin{equation}
    \label{eqn:reystress}
    T_{\rm{xy,Rey}} = \rho v_{\rm{x}} v'_{\rm{y}}.
\end{equation}
Due to our inclusion of self-gravity, we will also utilize the gravitational stress: 
\begin{equation}
    \label{eqn:gravstress}
    T_{xy,grav} = \frac{1}{4\pi \rm{G}}\frac{\partial \Phi}{\partial x}\frac{\partial \Phi}{\partial y}.
\end{equation}

Fig.~\ref{fig:stresses} shows the evolution of the volume-integrated toroidal magnetic field, and Maxwell, Reynolds, and gravitational stresses for all of our simulations. We see that for all simulations besides the weakly magnetized, strongly self-gravitating simulation (which fragments at approximately 5 orbits), the toroidal field saturates by approximately 5 orbits.  

For our strongly magnetized ($\beta^{\rm{mid}}_{0} = 10^{2.5}$) simulations, the evolution of the Maxwell and Reynolds stresses are comparable in the strongly self-gravitating, weakly self-gravitating, and pure MHD cases. Additionally, the gravitational stresses in the strongly self-gravitating and weakly self-gravitating simulations are closest in strength for the strongly magnetized simulations. While there are occasional peaks in the gravitational stress for the strongly self-gravitating, strongly magnetized simulation, it still remains several orders of magnitude weaker than the Reynolds or Maxwell stress. In short, the strength of self-gravity has minimal impact on the turbulent stresses in our strongly magnetized simulations.

The pure MHD and weakly self-gravitating cases for the weakly magnetized simulations show similar behavior in both the Maxwell and Reynolds stresses. However, the strongly self-gravitating, weakly magnetized simulation develops much stronger Maxwell and Reynolds stresses than in the other two weakly magnetized cases. As in the strongly magnetized simulations, the strongly self-gravitating simulation develops a drastically stronger gravitational stress compared to the weakly self-gravitating simulation. 

Across both magnetization values, the gravitational stress is at least two orders of magnitude weaker than the Maxwell and Reynolds stresses in all self-gravitating simulations, up to the point of catastrophic fragmentation of the weakly magnetized, strongly self-gravitating disk. In fact, even as the weakly magnetized, strongly self-gravitating disk undergoes fragmentation, the gravitational stress remains many orders of magnitude weaker than the Maxwell and Reynolds stresses for that simulation. This indicates that most of the energy from self-gravity does not reside in the gravitational stress. 

For a given magnetization, the pure MHD and weakly self-gravitating simulations display similar evolution of the Maxwell stress. The surviving strongly self-gravitating simulation ($\beta^{\rm{mid}}_{0} = 10^{2.5}$) also converges with its weakly self-gravitating and pure MHD counterparts. Additionally, in the strongly magnetized simulations, the Reynolds stress is similar across all three cases, and the difference between the gravitational stress in the strongly and weakly self-gravitating simulations is only about an order of magnitude. However, the weakly magnetized, strongly self-gravitating simulations show significantly enhanced Maxwell and Reynolds stresses (indicating significantly enhanced magnetic and fluid turbulence) compared to the  corresponding pure MHD and weakly self-gravitating simulations. 

In all six simulations, we see the characteristic linear growth of the MRI. In all simulations besides the weakly magnetized, strongly self-gravitating simulation, this iss followed by a plateau in the Maxwell and Reynolds stresses. In the weakly magnetized, strongly self-gravitating simulation (which ultimately fragmented), the Maxwell and Reynolds stresses begin the characteristic turnover at a similar point as the corresponding weakly self-gravitating and pure MHD simulations, but ultimately fail to saturate. We attribute this to the fragmentation of the disk.

\begin{deluxetable}{llll}
    \tablewidth{0pt}
    \tablecaption{MRI Diagnostics: Time-window averages of $\frac{T_{\rm{Max}}}{T_{\rm{Rey}}}$; the average for the strongly magnetized, pure MHD simulation was calculated excluding three data points from a drop in the Reynolds stress that lasted less than $0.1$ orbit. \label{tab:diagnostics}}
    \tablehead{
    \colhead{$\beta^{\rm{mid}}_{0}$} & \colhead{$Q_{0}$} & \colhead{$\Delta t_{\rm{avg}}$ [orbits]} & \colhead{$\left\langle \frac{T_{\rm{Max}}}{T_{\rm{Rey}}} \right\rangle_{t}$}
    }
\startdata
$10^{2.5}$ & 1.0 & 5.0 - 9.0 & 4.42  \\
$10^{2.5}$ & 10.0 & 5.0 - 15.0 & 4.45  \\
$10^{2.5}$ & MHD & 5.0 - 25.0 & 5.35  \\
$10^{4}$ & 1.0 & 5.0 - 6.0 & 3.71 \\
$10^{4}$ & 10.0 & 10.0 - 25.0 & 5.45  \\
$10^{4}$ & MHD & 10.0 - 25.0 & 5.41 \\
\enddata
\end{deluxetable}

\subsection{MRI Diagnostic: Ratio of Maxwell to Reynolds Stresses}\label{subsec:ratio}

\cite{Pessah06} demonstrated that across numerous numerical studies, the ratio of the saturated Maxwell stress to the saturated Reynolds stress faithfully adheres to $\rm{T_{Max}}/T_{\rm{Rey}} \approx 4$, regardless of initial computational set up. This ratio thus serves as a reliable diagnostic for the presence of MRI, and more specifically, the presence of the usual turbulent behavior driven by MRI. 

Table \ref{tab:diagnostics} shows the time-averaged $\rm{T_{Max}}/T_{\rm{Rey}}$ values for each simulation, as well as the time interval over which that average was performed. The averages start at 5 orbits, as all simulations developed approximate saturation of the toroidal field by that time. As noted in the table caption, the strongly magnetized, pure MHD simulation experienced a very brief drop in the Reynolds stress (see Fig.~\ref{fig:stresses}, at approximately 23 orbits). The three data points corresponding to that drop were excluded from the average, due to the extremely short duration and otherwise minimal variation in the Reynolds stress for that simulation. 

\begin{figure}
\includegraphics[width=8cm]{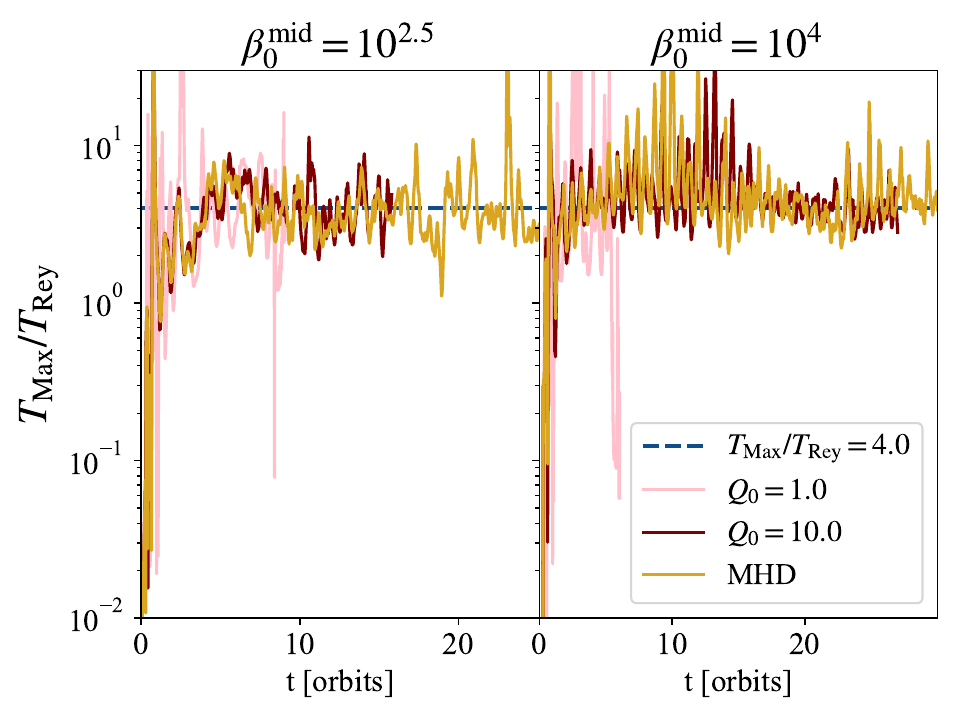}
\caption{Evolution of the ratio of the Maxwell to Reynolds stress for all simulations. Besides the weakly magnetized, strongly self-gravitating ($\beta^{\rm{mid}}_{0} = 10^{4}, Q_{0} = 1.0$) simulation, all simulations display oscillations around the canonical value for MRI of $T_{\rm{Max}}/T_{\rm{Rey}} \approx 4$.}
\label{fig:ratio}
\end{figure}

We see that all of the simulations (even the weakly magnetized, strongly self-gravitating simulation that ultimately fragments) adhere to the canonical $\rm{T_{Max}}/T_{\rm{Rey}} \approx 4$ ratio in saturation. Fig.~\ref{fig:ratio} shows the time evolution of the ratio of the Maxwell to Reynolds stress for each simulation. While clearly noisy, all simulations oscillate around approximately the expected value of $\rm{T_{Max}}/T_{\rm{Rey}} \approx 4$. It is worth noting in the final orbit of the weakly magnetized, strongly self-gravitating simulation, $\rm{T_{Max}}/T_{\rm{Rey}}$ displays greater amplitudes in its oscillations, which corresponds with the extreme growth of its gravitational stress and associated fragmentation. As the weakly magnetized simulations display greater noise than the strongly magnetized simulations, we propose that the noise is due to the weaker resolution of the MRI in the weakly magnetized case. 

\subsection{MRI Diagnostic: $\alpha_{\rm{B}}$-$\beta$ Relation}\label{subsec:alphamag}

Another commonly used diagnostic for the presence of MRI is the scaling relation of initial midplane magnetization, $\beta^{\rm{mid}}_{0}$, with the magnetic $\alpha_{\rm{B}}$ parameter, defined as: 
\begin{equation}
    \label{eqn:alphamag}
    \alpha_{\rm{B}} = \frac{T_{\rm{xy,Max}}+T_{\rm{xy,Rey}}}{p_{\rm{B}}}
\end{equation}
where $p_{\rm{B}}$ is the magnetic pressure. Various studies (e.g. 
\citealt{Hawley1995}, \citealt{Blackman2008}, \citealt{Salvesen2016a}) have found that in disks that have developed MRI, $\alpha_{\rm{B}}$ is essentially independent of the initial midplane magnetization, $\beta^{\rm{mid}}_{0}$. 

\begin{figure}
\includegraphics[width=8cm]{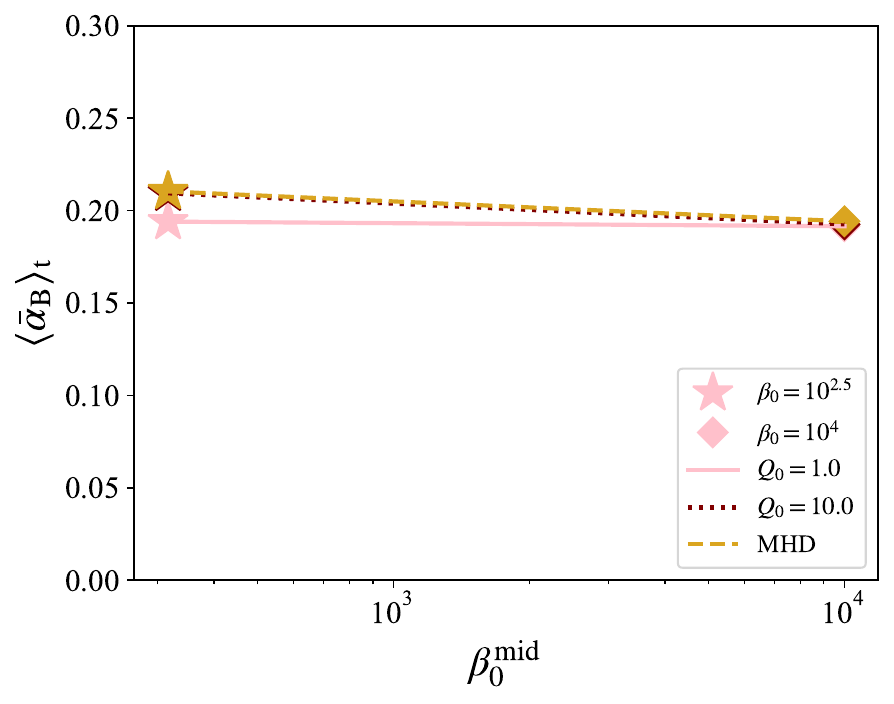}
\caption{Time-averaged, volume-averaged $\alpha_{\rm{B}}$ vs.~initial midplane magnetization, $\beta^{\rm{mid}}_{0}$, as described in Eq.~\ref{eqn:alphamag}. $\alpha_{\rm{B}}$ is essentially independent of $\beta^{\rm{mid}}_{0}$, providing support for the existence of MRI in our simulations. Marker style corresponds to initial magnetization (stars for $\beta^{\rm{mid}}_{0} = 10^{2.5}$, diamonds for $\beta^{\rm{mid}}_{0} = 10^{4}$), line colors correspond to strength of self-gravity. The time average for $\alpha_{\rm{B}}$ for our weakly magnetized, strongly self-gravitating simulation (which ultimately fragments) was only performed over the first 5 orbits of the simulation, due to the fragmentation.}
\label{fig:alphamagbeta}
\end{figure}

Fig. \ref{fig:alphamagbeta} shows the $\langle \langle \alpha_{\rm{B}} \rangle_{\rm{V}} \rangle_{\rm{t}}$ values for our simulations against the corresponding initial magnetization. The time averages were performed over the duration of the simulations, with the exception of our weakly magnetized, strongly self-gravitating simulation. Due to the ultimate fragmentation of that simulation, the time average was performed over the first 5 orbits.\footnote{Since the averaging interval 0--5 orbits includes the linear growth of MRI, we repeated the average starting from 4 orbits, where the toroidal field turns over. This gives a virtually identical result for $\alpha_{\rm B}$.} While we understand the limitations of fitting a line to two data points, we still believe doing so provides not only insight as to whether the MRI is present, but also bears on whether the MRI may have developed differently with strength of self-gravity. We find that our strongly self-gravitating, weakly self-gravitating, and pure MHD simulations aqll display the characteristic independence of $\alpha_{\rm{B}}$ and $\beta^{\rm{mid}}_{0}$.  

From the results of Sections \ref{subsec:stresses}--\ref{subsec:alphamag}, we conclude that the MRI develops normally in all five of our surviving simulations, and thus the abnormalities in our simulation that fragments are due to the fragmentation.  

\subsection{Disk Magnetization}\label{subsec:midplanebeta}

Having established the presence of normal MRI in our simulations that do not fragment, we now will establish the presence of magnetic pressure dominance in our strongly magnetized ($\beta^{\rm{mid}}_{0} = 10^{2.5}$) disks, and its absence in our weakly magnetized ($\beta^{\rm mid}_{0} = 10^{4}$) disks. Magnetic pressure dominance arises due to magnetic pressure exceeding gas pressure within the disk \citep{BegelmanPringle2007}. Pure MHD studies (e.g. \citealt{Salvesen2016a}) find that $\beta^{\rm{mid}}_{0} \lesssim 10^{3}$ is sufficient for a disk to develop magnetic pressure dominance. Here, we present vertical density profiles of the magnetization of our simulations, which we define as the time average (over the time interval stated in Table \ref{tab:diagnostics}) of the horizontal average of magnetization at each $z$. 

\begin{figure}
\includegraphics[width=8cm]{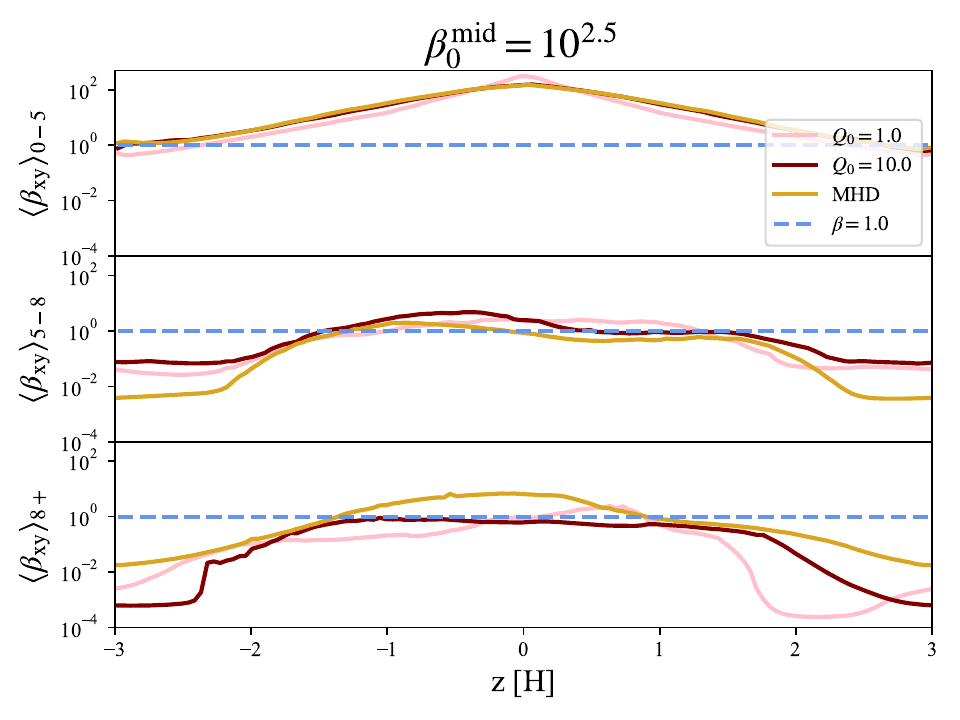}
\caption{Time window-averaged profiles of magnetization for our strongly magnetized ($\beta^{\rm mid}_{0} = 10^{2.5}$) simulations. The subscript on the $y$-axis labels indicates orbits over which the average was performed, with the bottom panel representing an average performed from the eighth orbit until the end of the simulation. Shortly after the plateauing of the Maxwell stress (see Fig.~\ref{fig:stresses}), nearly the entire disk becomes magnetically dominated for all three simulations.}
\label{fig:betaprofile_b25}
\end{figure}

Fig.~\ref{fig:betaprofile_b25} shows time window-averaged magnetization profiles for our strongly magnetized simulations, showing that essentially the entire disk becomes magnetically dominated shortly after the Maxwell stress plateaus (see Fig.~\ref{fig:stresses}). Interestingly, the pure MHD case does not develop stronger magnetic domination than the self-gravitating simulations. The bottom panel represents an average performed over times for which the strongly self-gravitating simulation (initialized to $Q_{0} = 1.0$) had stabilized and developed $Q' > 1.0$ (see Sec.~\ref{sec:analysis}). Once stabilized, its magnetization profile is very similar to that of the weakly self-gravitating simulation (initialized to $Q_{0} = 10.0$), despite having significantly different $Q'$ values. Perhaps most surprising is that despite the strongly self-gravitating simulation developing slightly higher density in the midplane (see Fig.~\ref{fig:densityprofiles}), its magnetization profile is not meaningfully different from that of the weakly self-gravitating case and its magnetization is slightly stronger than the pure MHD case. In Fig.~\ref{fig:stresses}, we can see that the toroidal field for the strongly self-gravitating case is slightly stronger than the other two cases. This slight enhancement is likely responsible for the slightly stronger magnetization of the strongly self-gravitating case, though we leave exploration of this minor enhancement to future work.

\begin{figure}
\includegraphics[width=8cm]{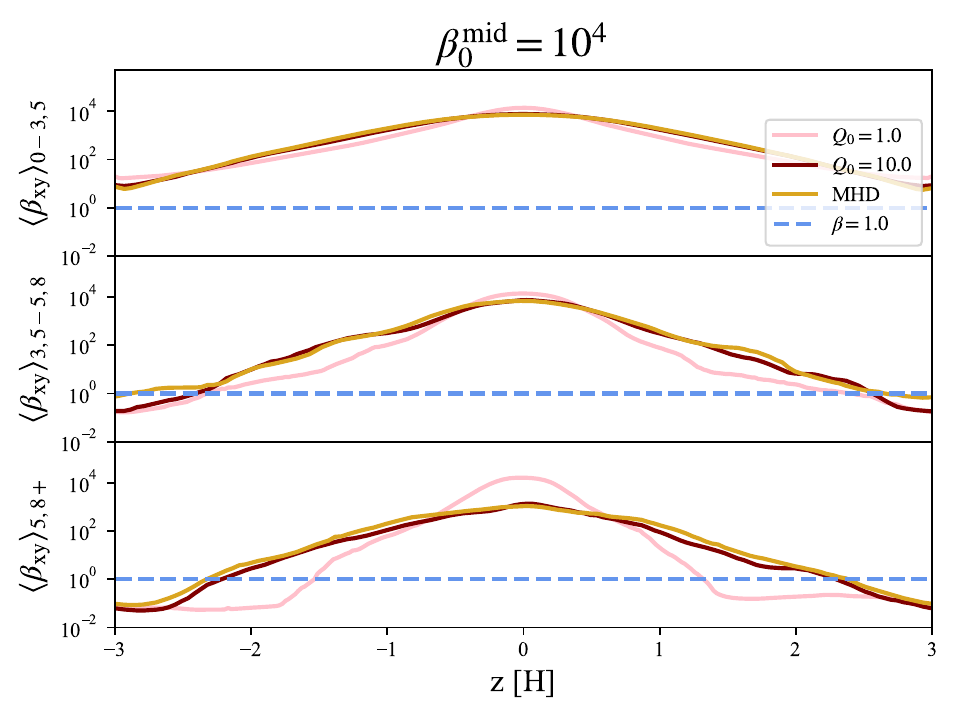}
\caption{Time window-averaged profiles of magnetization for our weakly magnetized ($\beta^{\rm{mid}}_{0} = 10^{4}$) simulations. As in Fig.~\ref{fig:betaprofile_b25}, the subscripts on the $y$-axis labels correspond to the orbits over which the average was performed; however, as our weakly magnetized, strongly self-gravitating simulation fragmented quickly, the time-windows were adjusted, represented by the lower values in each subscript (e.g., first panel is averaged over the first three orbits, instead of first five, for the $Q_{0} = 1.0$ simulation). In contrast to the strongly magnetized simulations, none of the weakly magnetized simulations develop magnetic dominance in any region besides near the vertical boundaries.}
\label{fig:betaprofile_b4}
\end{figure}

In contrast, Fig.~\ref{fig:betaprofile_b4} shows that none of our weakly magnetized simulations develop regions of magnetic domination within $|z| \leq 2H$. The weakly self-gravitating and pure MHD simulations largely evolve similarly, while the strongly self-gravitating simulation (which ultimately fragmented) develops significantly weaker magnetization in the midplane prior to fragmentation. 

These results show that all three of our strongly magnetized simulations develop magnetic dominance essentially throughout the disk. Meanwhile, none of our weakly magnetized simulations do the same. 

\subsection{Vertical Gas Density Profile}\label{subsec:profile}

\begin{figure}
\includegraphics[width=8cm]{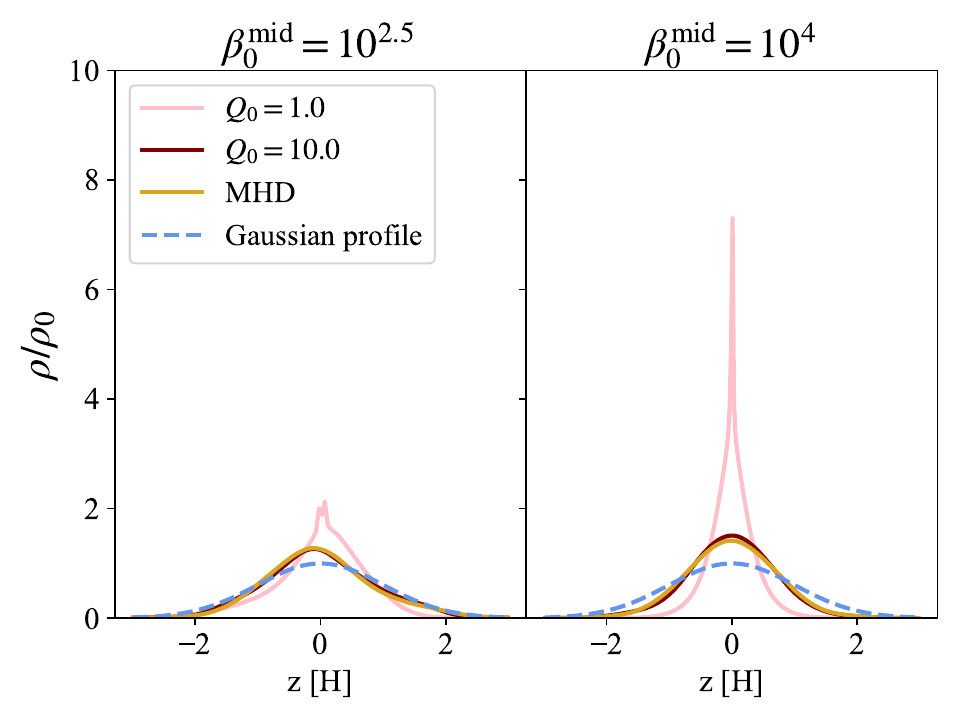}
\caption{Time-averaged vertical density profiles for all six simulations (over the same duration as the averages in Table~\ref{tab:diagnostics}). In both strongly and weakly magnetized simulations, the weakly self-gravitating and pure MHD disks have almost identical profiles. However, the strongly magnetized profiles are noticeably wider and shorter than the weakly magnetized ones. For both strongly and weakly magnetized disks, the strongly self-gravitating simulations develop significantly higher peaks in the density profile. However, the strongly magnetized, strongly self-gravitating profile has a peak only approximately half that of the weakly magnetized case (and significantly wider).}
\label{fig:densityprofiles}
\end{figure}

Having established that our strongly magnetized disks become magnetically dominated, while our weakly magnetized disks do not, we will now discuss how magnetic pressure dominance (or lack thereof) impacts our disks' structure. In the absence of other physics, self-gravity will contract a disk vertically, resulting in a narrower, more strongly peaked vertical density profile. 
Fig.~\ref{fig:densityprofiles} shows the horizontally averaged density as a function of $z$ for all simulations, time-averaged over the same time windows as listed in Table~\ref{tab:diagnostics}. 

\begin{figure*}
\centering
\includegraphics[width=18cm]{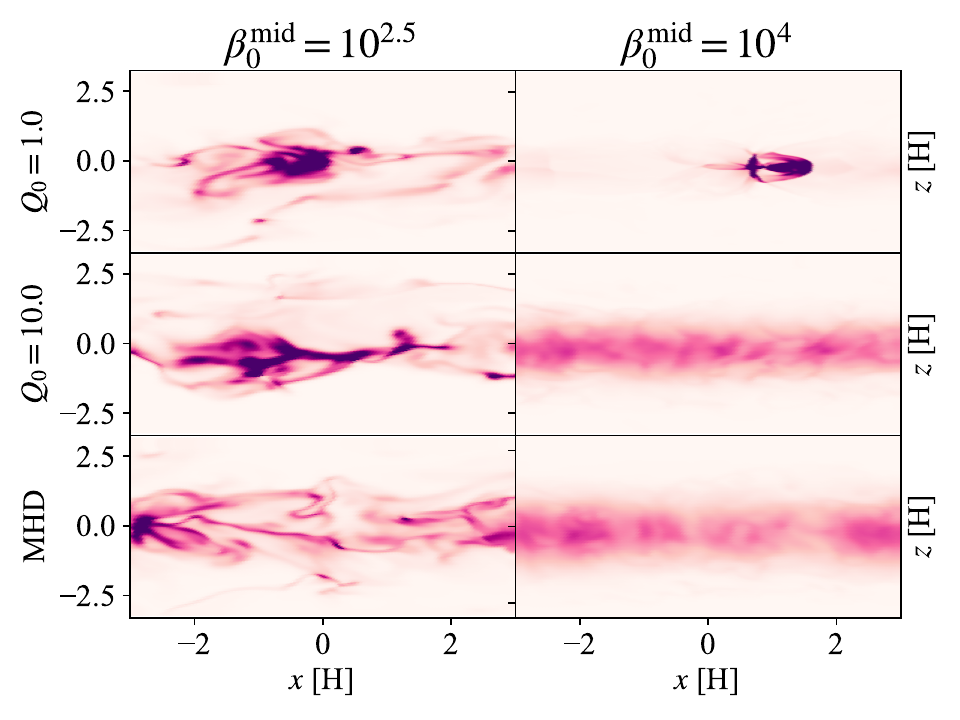}
\caption{Gas density of all 6 simulations in their final orbits, sliced at $y = 0.0$. The strongly magnetized ($\beta^{\rm{mid}}_{0} = 10^{2.5}$) simulations are much more turbulent and disordered than their weakly magnetized ($\beta^{\rm{mid}}_{0} = 10^{4}$) counterparts. While the strongly magnetized, strongly self-gravitating disk develops strong density maxima at the midplane, the corresponding weakly self-gravitating disk shows some density peaks off midplane.}
\label{fig:slices}
\end{figure*}

All three strongly magnetized ($\beta^{\rm{mid}}_{0} = 10^{2.5}$) simulations develop extremely similar density profiles. The strongly self-gravitating simulation develops a small bump around the midplane, due to the formation of high density clumps (see Fig.~\ref{fig:slices} for an example of such an object). They all develop a slightly narrower and higher peak than the Gaussian profile they were initialized to, as shown in the left panel of Fig.~\ref{fig:densityprofiles}. The strength of self-gravity correlates with the narrowness and height of the profile's peak, though fairly weakly. 

In contrast, the pure MHD weakly magnetized ($\beta^{\rm{mid}}_{0} = 10^{4}$) simulation develops a notably narrower and taller peak than the corresponding strongly magnetized simulations, indicating that the superior magnetic pressure of the strongly magnetized simulations plays a primary role in their disk structure. The weakly self-gravitating, weakly magnetized simulation develops a nearly identical profile to the pure MHD case, though the strongly self-gravitating counterpart develops an extremely narrow and high peak. All three weakly magnetized disks develop narrower profiles than the corresponding strongly magnetized disks.  

In Sections \ref{subsec:stresses}--\ref{subsec:alphamag}, we showed that the MRI develops normally in our five simulations that don't ultimately fragment. In Sections \ref{subsec:midplanebeta} and \ref{subsec:profile}, we showed the strongly magnetized disks become magnetically dominated throughout and subsequently retain a thicker vertical density profile, while the weakly magnetized disks do not. 

\section{Analysis}\label{sec:analysis}

In this section, we will show that after becoming magnetically dominated as shown in Sec.~\ref{sec:results}, the strongly magnetized, strongly self-gravitating simulation ultimately survives and stabilizes against gravitational instability, indicating that a magnetic pressure-dominated disk can likely avoid catastrophic fragmentation. We will also show that our weakly magnetized, strongly self-gravitating simulation (which did not become magnetically dominated) fails to stabilize. Finally, we will show that even in non-negligibly self-gravitating regions of disks that are robust against GI, a weakly magnetized disk becomes less stable, while a magnetic pressure dominated, strongly magnetized disk becomes more stable. 

Fig.~\ref{fig:slices} shows $x-z$ slices of gas density, taken at $y = 1.0$, for the last time steps from each simulation. Our strongly magnetized simulations are significantly more turbulent than our weakly magnetized simulations. For both magnetizations, the strength of self-gravity increases the inhomogeneity of the disk. All three strongly magnetized disks show similar structure to each other, with regions of higher density connected by filaments of lower density. The pure MHD and weakly self-gravitating, weakly magnetized simulations are very similar, with the latter having marginally more inhomogeneity than the former. However, the strongly self-gravitating, weakly magnetized disk shows visible fragmentation of the disk. 

\subsection{Disk Stability}\label{subsec:qevolution}

\begin{figure}[t]
\includegraphics[width=9cm]{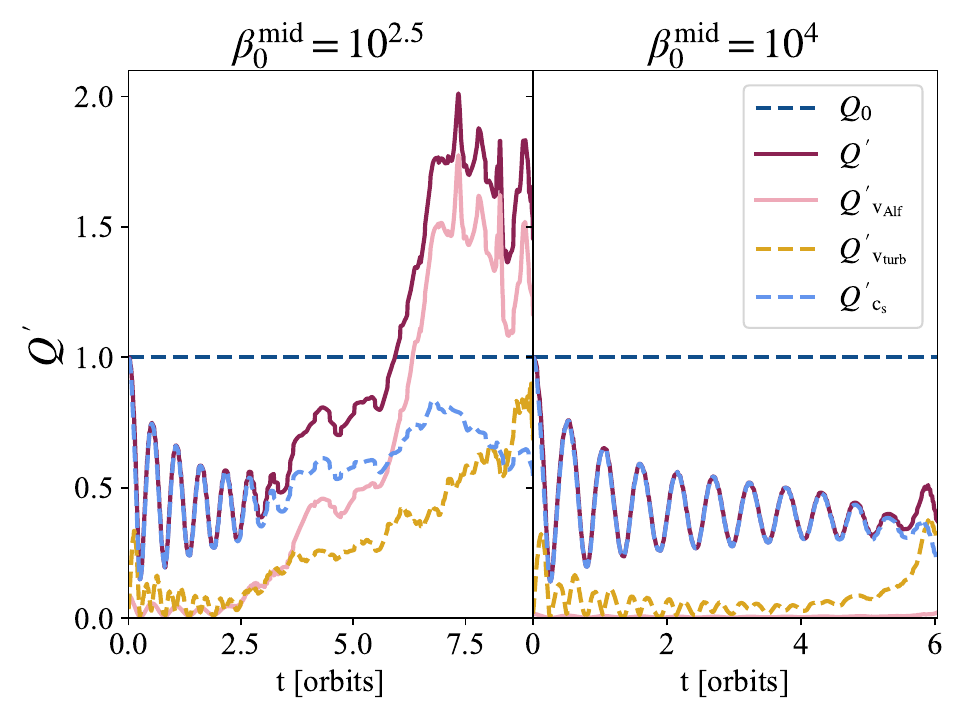}
\caption{Evolution of the modified Q-parameter (Eq.~\ref{eqn:modQ}) and its components for strongly self-gravitating simulations; left panel corresponds to the strongly magnetized case ($\beta^{\rm{mid}}_{0} = 10^{2.5}$) and right panel to the weakly magnetized case ($\beta^{\rm{mid}}_{0} = 10^{4}$).}
\label{fig:q1evolve}
\end{figure}

\begin{figure}[t]
\includegraphics[width=9cm]{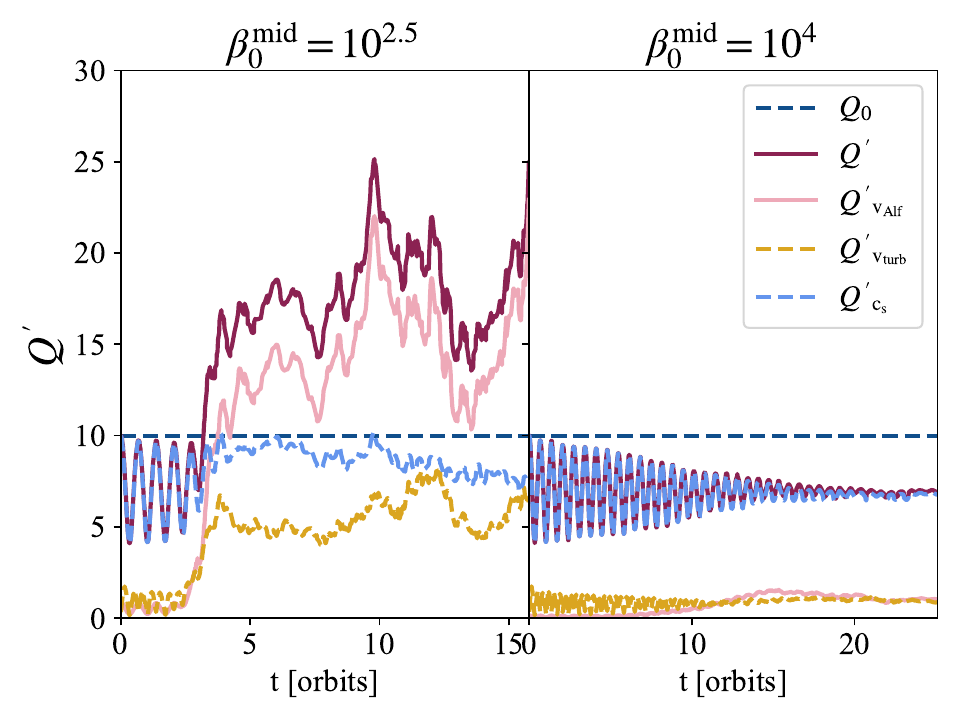}
\caption{Same as Fig.~\ref{fig:q1evolve}, for the weakly self-gravitating simulations; note the different scale on the $y$-axis.}
\label{fig:q10evolve}
\end{figure}

When attempting to answer any question of disk stability against GI, the most important parameter is the Toomre parameter. In Sec.~\ref{subsec:modQ}, we explained our modified definition of the Toomre parameter, to include terms from additional important sources of pressure in the disk besides the gas pressure. We will now discuss the evolution of $Q'$ and its implications for disk stability. In Figures \ref{fig:q1evolve} and \ref{fig:q10evolve}, each component of $Q'$ is given by $Q'_{\rm{v_i}} = \frac{\Omega \rm{v_{i}}}{\pi G \Sigma_{0}}$, where $v_{\rm{i}} = c_{\rm{s}}, v_{\rm{A}}, v_{\rm{turb}}$. 

As shown in Fig.~\ref{fig:q1evolve}, both of our strongly self-gravitating disks undergo a transient phase of drastic reduction in $Q'$ (Eq.~\ref{eqn:modQ}), indicating the onset of GI. However, our strongly magnetized disk stabilizes at the end of the MRI's linear growth phase (the turnover in the Maxwell stress in Fig.~\ref{fig:stresses}). Note that $Q'$ starts to grow in our strongly magnetized simulations at the same point where the toroidal field saturates. However, the weakly magnetized $Q_{0} = 1.0$ disk fails to stabilize and ultimately fragments, as defined in Sec.~\ref{subsec:defs}. While the weakly magnetized case does briefly show a slight growth in $Q'$, driven by the contribution from the fluid turbulence, it is short-lived.

In our strongly magnetized, strongly self-gravitating simulation, the dominant contribution to $Q'$ is that from  magnetic pressure. In contrast to the strongly magnetized case, the evolution of $Q'$ in the weakly magnetized simulation is always dominated by the gas pressure: no other term is ever important. It is thus clear that strong magnetic pressure is necessary for stabilization of critically stable disks against GI, given not only these trends in $Q'$ evolution but also the results of Sections \ref{subsec:midplanebeta} and \ref{subsec:profile}. This is also in agreement with the results of \cite{tsung2025doesmagneticfieldpromote}. 

Fig.~\ref{fig:q10evolve} shows the evolution of $Q'$ for the weakly self-gravitating simulations. In the strongly magnetized simulations, $Q'$ grows beyond its initial value. In contrast, as in the strongly self-gravitating case, $Q'$ of the weakly magnetized disk never recovers to its initial value although the disk remains robustly stable against GI.  We find that $Q'$ of the strongly magnetized, weakly self-gravitating simulation is also dominated by the contribution from magnetic pressure. In fact, magnetic pressure is even more dominant in the strongly magnetized, weakly self-gravitating case than in the corresponding strongly self-gravitating case. This indicates that while the fluid turbulence contribution doesn't rise to the point of even co-dominance, it is still enhanced in the strongly self-gravitating, strongly magnetized case compared to the weakly self-gravitating case. 

The weakly magnetized, weakly self-gravitating disk follows a similar behavior to its strongly self-gravitating counterpart: $Q'$ is dominated by the contribution from the gas pressure, and ends up decreasing compared to its initial value. In the weakly self-gravitating case, the weakly magnetized disk reaches a steady state still robustly stable against GI. However, this reduction in stability indicates that weakly magnetized disks may fragment in regions of the disk thought to be stable against GI from a simple hydrodynamical analysis.  

Interestingly, there is no simulation where the traditional definition of the Toomre parameter (see Sec.~\ref{subsec:modQ}) increases from its initial value. Even in the presence of minor inhomogeneity --- which develops even in pure MHD \citep[e.g.][]{Salvesen2016a} --- the traditional definition of $Q$ no longer adequately captures the stability of the disk against GI. This is due to the traditional definition considering only hydrodynamical effects, which balance gas pressure against self-gravity. When MHD effects are included, inhomogeneity can develop from magnetic effects alone. On the one hand, this results in a clumpier disk even if fluctuations in the gas pressure are weak. On the other hand, however, MRI-driven turbulence increases the effective pressure, providing a stabilizing influence against GI. 

All of our simulations undergo a period of destabilization, as shown in both Fig.~\ref{fig:q1evolve} and Fig.~\ref{fig:q10evolve}. Our strongly magnetized, strongly self-gravitating simulation maintains $Q' < 1.0$ for over 5 orbits, and even falls to $Q' \approx 0.2$ several times in its first several orbits. Despite this, both of our self-gravitating strongly magnetized disks ultimately demonstrate enhanced stability beyond their initialization. This shows that magnetic pressure dominance, driven by MRI, can stabilize a critically stable disk against GI, even when the disk initially collapses to significantly below $Q' = 1.0$. We leave study of strongly magnetized disks initialized to $Q_{0} < 1.0$ to future work.

In contrast, both of our self-gravitating, weakly magnetized simulations failed to recover their initial level of stability.  In Figures \ref{fig:q1evolve} and \ref{fig:q10evolve}, this failure is manifestly due to the negligible contributions from magnetic pressure and turbulent pressure. Even in the absence of disk breathing, $Q_{0} = 1.0$ is not a stable equilibrium, and minor density inhomogeneity would cause $Q'$ to decrease. With negligible contributions to $Q'$ from all terms besides the gas pressure, recovery to $Q' > 1.0$ would be impossible. Fig.~\ref{fig:q10evolve} shows that even once MRI is established and disk breathing damps out, the magnetic pressure and turbulent pressure produced by $\beta^{\rm{mid}}_{0} = 10^{4}$ are much too weak to contribute meaningfully to the stability of the disk. For our weakly magnetized, weakly self-gravitating simulations, disk breathing had mostly damped out by approximately 15 orbits; despite that, $Q'$ did not grow at all. Thus, it is clear that $\beta^{\rm{mid}}_{0} = 10^{4}$ is insufficient to lead to enhanced stability, even in the weakly self-gravitating regime. Furthermore, the inhomogeneity innately introduced by MRI in a stratified disk (e.g. \citealt{Salvesen2016a}) may cause an even weakly self-gravitating, $\beta^{\rm{mid}}_{0} = 10^{4}$ disk to become less stable against GI.  

\section{Discussion}\label{sec:disussion}
The most striking result from this study is that the MRI can successfully suppress fragmentation in an isothermal disk, if adequate resolution is used to resolve the most unstable modes. For $\beta^{\rm{mid}}_{0} = 10^{2.5}$, our MRI quality factors are sufficient to resolve the fastest growing mode of the MRI. For our weakest magnetized case, our MRI quality factor is still higher than that employed in other studies. We find that magnetic pressure dominance plays the defining role in disk stabilization. Furthermore, we find that if there is sufficient magnetization to drive magnetic pressure dominance, the disk behaves as it does in pure MHD. While we initialized our strongly self-gravitating simulations to $Q_{0} = 1.0$, denoting critical stability against GI, those disks dropped to $Q' < 1.0$ immediately. The strongly magnetized disk, which stabilized after 5 orbits, fell to as low as $Q' \approx 0.2$ multiple times. This indicates that magnetic pressure dominance not only can stabilize a critically stable disk, but also a disk that undergoes periods of severe instability. 

 Our strongly magnetized disk that survived initialization to $Q_{0} = 1.0$ has sufficiently strong magnetic fields that buoyancy may lead to some escape of the toroidal field. However, recent work suggests that due to the high magnetization in the midplane developed by our strongly magnetized simulations ($\beta \lesssim 1.0$), the toroidal field is unlikely to escape \citep{Squire2025Rapid}. We leave more thorough study of the relationship between magnetization and accretion rate through the self-gravitating regions of AGN disks to future work. 

Other mechanisms, not considered in this paper, 
may also contribute to stabilizing the outer regions of AGN disks against GI. \cite{Chen_2023} showed that radiation alone cannot provide enough pressure to stabilize a disk against GI, and therefore it is highly unlikely that there exists a parameter space in which radiation pressure is the dominant factor in stabilizing a quasar disk. However, the coupling of radiation to other physical processes may expand the magnetization parameter space that permits disk survival. In threshold cases, where the magnetization is insufficient to drive strong magnetic pressure dominance but is sufficient to exert significant magnetic pressure against the contraction of self-gravity, additional pressure from radiation may contribute enough to stabilize the disk. 
Massive stars (or their remnants), embedded in the disk, might inject enough energy into the disk to contribute to such stabilization (e.g. \citealt{Syer91, Pawel93, Huang23, EpsteinMartin25}).

Further, large scale magnetic torques in the disk can increase the inflow speed of matter through the disk \citep{BegelmanSilk23}. In turn, the density of the disk would decrease, reducing the impact of self-gravity. In the extreme limit of the magnetic torques model, some recent work has shown that an extremely strong toroidal field advected from the nuclear environment can stabilize the outer regions of AGN disks \citep{Hopkins2024FORGEda, Hopkins2024FORGEdb, Hopkins2024FORGEdc}. Any of these processes would potentially extend the region of stability further outwards than a basic hydrodynamic or magnetohydrodynamic calculation would indicate. 

AGN disks are intrinsically multi-physics systems. While we have shown that magnetic pressure dominance alone can stabilize an AGN disk against GI, hybrid means of disk stabilization remains a promising avenue of exploration. 
Further study of this topic may lend insight into how different disk stabilization mechanisms may be impacted by environmental conditions, as well as how different disk stabilization mechanisms may influence the growth and evolution of quasar disks, and therefore the quasars themselves. 

\section{Conclusions}\label{sec:conclude}

In this work, we have performed simulations to determine whether or not an isothermal, magnetized, self-gravitating disk can stabilize against the GI and avoid catastrophic fragmentation, even if initialized to critical stability ($Q_{0} = 1.0$). We found that our strongly magnetized disks ($\beta^{\rm{mid}}_{0} = 10^{2.5}$) successfully stabilized, despite both disks initially undergoing a short period of severe destabilization, while our weakly magnetized disk ($\beta^{\rm{mid}}_{0} = 10^{4}$) did not stabilize against fragmentation. The strongly magnetized, strongly self-gravitating disk that stabilized did so rapidly, reaching $Q' > 1.0$ in slightly over 5 orbits (concurrent with the turnover in the Maxwell stress growth), where $Q'$ is the modified Toomre parameter defined in Sec.~\ref{subsec:modQ} to include additional sources of pressure besides gas pressure. 

We also performed simulations with the same pair of magnetization values, but initialized to $Q_{0} = 10.0$, which corresponds to non-negligible self-gravity but robust against the GI. Both weakly self-gravitating disks remained stable. However, only the strongly magnetized disk displayed enhanced stability, while the weakly magnetized disk became less stable. In both the strongly and weakly self-gravitating, strongly magnetized simulations (i.e. the simulations that developed enhanced stability against GI), the contribution from magnetic pressure was the dominant contribution to disk stabilization.

Our strongly magnetized simulations developed magnetic pressure dominance throughout nearly the entire disk within 8 orbits. Our results suggest that magnetic pressure dominance plays a critical, and perhaps necessary role in the stabilization of the outer regions of AGN disks against GI. 

\indent Our most important findings can be summarized as follows: 

\begin{itemize}
    \item An initial midplane magnetization of $\beta^{\rm{mid}}_{0} = 10^{2.5}$ is sufficient to stabilize a critically stable disk ($Q_{0} = 1.0$) against GI, even if the disk undergoes a transient period of severe instability (as low as $Q' \approx 0.2$).
    \item An initial midplane magnetization of $\beta^{\rm{mid}}_{0} = 10^{4}$ is not sufficient to stabilize a critically stable disk. 
    \item An initial midplane magnetization of $\beta^{\rm{mid}}_{0} = 10^{2.5}$ will enhance the stability of a robustly stable, self-gravitating disk ($Q_{0} = 10.0$), while $\beta^{\rm{mid}}_{0} = 10^{4}$ becomes less stable. However, the latter will not fall to the unstable regime.  
    \item Stabilization of disks with $\beta^{\rm{mid}}_{0} = 10^{2.5}$ occurs rapidly, concurrent with the establishment of the toroidal magnetic field generated by MRI; additionally, stability enhancement is driven predominantly by magnetic pressure.
    \item The strength of self-gravity does not prevent magnetic pressure dominance throughout (nearly) the entire disk for $\beta^{\rm{mid}}_{0} = 10^{2.5}$.
\end{itemize}

\section*{Acknowledgements}
HGD would like to thank William D. Dorland for productive conversations. HGD acknowledges support from NASA FINESST Fellowship 80NSSC22K1753. MCB and PJA acknowledge support from NASA Astrophysics Theory Program grant 80NSSC24K0940. The results presented are from simulations performed using resources provided by the NASA High-End Computing (HEC) Program through the NASA Advanced Supercomputing (NAS) Division at Ames Research Center. 

\section*{Data Availability}

The data used for this study is available upon reasonable request to the authors.

\bibliography{bibliography}{}

\begin{thebibliography}{}
\expandafter\ifx\csname natexlab\endcsname\relax\def\natexlab#1{#1}\fi
\providecommand{\url}[1]{\href{#1}{#1}}
\providecommand{\dodoi}[1]{doi:~\href{http://doi.org/#1}{\nolinkurl{#1}}}
\providecommand{\doeprint}[1]{\href{http://ascl.net/#1}{\nolinkurl{http://ascl.net/#1}}}
\providecommand{\doarXiv}[1]{\href{https://arxiv.org/abs/#1}{\nolinkurl{https://arxiv.org/abs/#1}}}

\bibitem[{P. Armitage(2011)Armitage}]{Armitage2011}
Armitage, P. 2011, \bibinfo{title}{Dynamics of Protoplanetary Disks,} Annual Review of Astronomy and Astrophysics, 49, p. 195

\bibitem[{X. Bai \& J.~M. Stone(2013)Bai \& Stone}]{BaiStone2013}
Bai, X., \& Stone, J.~M. 2013, \bibinfo{title}{Local Study of Accretion Disks with a Strong Vertical Magnetic Field: Magnetorotational Instability and Disk Outflow,} The Astrophysical Journal, 767, p. 30

\bibitem[{S.~A. Balbus \& J.~F. Hawley(1991)Balbus \& Hawley}]{BalbusHawley91}
Balbus, S.~A., \& Hawley, J.~F. 1991, \bibinfo{title}{A Powerful Local Shear Instability in Weakly Magnetized Disks I. Linear Analysis,} The Astrophysical Journal, 376, p. 214

\bibitem[{M.~C. Begelman \& J.~E. Pringle(2007)Begelman \& Pringle}]{BegelmanPringle2007}
Begelman, M.~C., \& Pringle, J.~E. 2007, \bibinfo{title}{Accretion discs with strong toroidal magnetic fields,} MNRAS, 375, p. 1070

\bibitem[{M.~C. Begelman \& J. Silk(2023)Begelman \& Silk}]{BegelmanSilk23}
Begelman, M.~C., \& Silk, J. 2023, \bibinfo{title}{Magnetic fields catalyse massive black hole formation and growth,} Monthly Notices of the Royal Astronomical Society: Letters, 526, L94, \dodoi{10.1093/mnrasl/slad124}

\bibitem[{E. Blackman {et~al.}(2008)Blackman, Penna, \& Varnière}]{Blackman2008}
Blackman, E., Penna, R., \& Varnière, P. 2008, \bibinfo{title}{Empirical relation between angular momentum transport and thermal-to-magnetic pressure ratio in shearing box simulations,} New Astronomy, 13, 244, \dodoi{https://doi.org/10.1016/j.newast.2007.10.004}

\bibitem[{R.~D. Blandford \& D.~G. Payne(1982)Blandford \& Payne}]{blandfordpayne}
Blandford, R.~D., \& Payne, D.~G. 1982, \bibinfo{title}{Hydromagnetic flows from accretion discs and the production of radio jets,} Monthly Notices of the Royal Astronomical Society, 199, 883, \dodoi{10.1093/mnras/199.4.883}

\bibitem[{D. Carrera {et~al.}(2015)Carrera, Johansen, \& Davies}]{Carrera2015}
Carrera, D., Johansen, A., \& Davies, M.~B. 2015, \bibinfo{title}{How to form planetesimals from mm-sized chondrules and chondrule aggregates,} Astronomy \& Astrophysics, 579

\bibitem[{Y.-X. Chen {et~al.}(2023)Chen, Jiang, Goodman, \& Ostriker}]{Chen_2023}
Chen, Y.-X., Jiang, Y.-F., Goodman, J., \& Ostriker, E.~C. 2023, \bibinfo{title}{3D Radiation Hydrodynamic Simulations of Gravitational Instability in AGN Accretion Disks: Effects of Radiation Pressure,} The Astrophysical Journal, 948, 120, \dodoi{10.3847/1538-4357/acc023}

\bibitem[{H. {Deng} {et~al.}(2021){Deng}, {Mayer}, \& {Helled}}]{Deng2012}
{Deng}, H., {Mayer}, L., \& {Helled}, R. 2021, \bibinfo{title}{{Formation of intermediate-mass planets via magnetically controlled disk fragmentation},} Nature Astronomy, 5, 440, \dodoi{10.1038/s41550-020-01297-6}

\bibitem[{A.~J. {Dittmann} \& M.~C. {Miller}(2020){Dittmann} \& {Miller}}]{Dittmann2020}
{Dittmann}, A.~J., \& {Miller}, M.~C. 2020, \bibinfo{title}{{Star formation in accretion discs and SMBH growth},} \mnras, 493, 3732, \dodoi{10.1093/mnras/staa463}

\bibitem[{M. Epstein-Martin {et~al.}(2025)Epstein-Martin, Tagawa, Haiman, \& Perna}]{EpsteinMartin25}
Epstein-Martin, M., Tagawa, H., Haiman, Z., \& Perna, R. 2025, \bibinfo{title}{Time-dependent models of AGN discs with radiation from embedded stellar-mass black holes,} Monthly Notices of the Royal Astronomical Society, 537, 3396, \dodoi{10.1093/mnras/staf237}

\bibitem[{J. Ferreira \& G. Pelletier(1993)Ferreira \& Pelletier}]{FerreiraPelletier93a}
Ferreira, J., \& Pelletier, G. 1993, \bibinfo{title}{Magnetized Accretion-Ejection Structures I. General Statements,} Astronomy \& Astrophysics, 276, p. 625

\bibitem[{S. Fromang {et~al.}(2004)Fromang, Balbus, Terquem, \& De~Villiers}]{Fromang2004}
Fromang, S., Balbus, S., Terquem, C., \& De~Villiers, J.~P. 2004, \bibinfo{title}{Evolution of Self-Gravitating Magnetized Disks II. Interaction Between Magnetohydrodynamic Turbulence and Gravitational Instabilities,} The Astrophysical Journal, 616, p. 364

\bibitem[{S. Fromang {et~al.}(2013)Fromang, Latter, Lesur, \& Ogilvie}]{Fromang2013}
Fromang, S., Latter, H.~N., Lesur, G., \& Ogilvie, G.~I. 2013, \bibinfo{title}{Local outflows from turbulent accretion disks,} Astronomy \& Astrophysics, 552

\bibitem[{E. Gaburov {et~al.}(2012)Gaburov, Johansen, \& Levin}]{Gaburov2012}
Gaburov, E., Johansen, A., \& Levin, Y. 2012, \bibinfo{title}{MAGNETICALLY LEVITATING ACCRETION DISKS AROUND SUPERMASSIVE BLACK HOLES,} The Astrophysical Journal, 758, 103, \dodoi{10.1088/0004-637X/758/2/103}

\bibitem[{C.~F. Gammie(2001)Gammie}]{Gammie2001}
Gammie, C.~F. 2001, \bibinfo{title}{Nonlinear Outcome of Gravitational Instability in Cooling, Gaseous Disks,} The Astrophysical Journal, 553, p. 174

\bibitem[{J. Goodman(2003)Goodman}]{Goodman2003}
Goodman, J. 2003, \bibinfo{title}{Self-gravity and quasi-stellar object discs,} Monthly Notices of the Royal Astronomical Society, 339, 937, \dodoi{10.1046/j.1365-8711.2003.06241.x}

\bibitem[{J.~F. {Hawley} {et~al.}(1995){Hawley}, {Gammie}, \& {Balbus}}]{Hawley1995}
{Hawley}, J.~F., {Gammie}, C.~F., \& {Balbus}, S.~A. 1995, \bibinfo{title}{{Local Three-dimensional Magnetohydrodynamic Simulations of Accretion Disks},} \apj, 440, 742, \dodoi{10.1086/175311}

\bibitem[{J.~F. Hawley {et~al.}(2011)Hawley, Guan, \& Krolik}]{Hawley2011}
Hawley, J.~F., Guan, X., \& Krolik, J.~H. 2011, \bibinfo{title}{ASSESSING QUANTITATIVE RESULTS IN ACCRETION SIMULATIONS: FROM LOCAL TO GLOBAL,} The Astrophysical Journal, 738, p. 84

\bibitem[{P.~F. Hopkins {et~al.}(2024{\natexlab{a}})Hopkins, Grudic, Kremer, Offner, Guszejnov, \& Rosen}]{Hopkins2024FORGEdc}
Hopkins, P.~F., Grudic, M.~Y., Kremer, K., {et~al.} 2024{\natexlab{a}}, \bibinfo{title}{FORGE\textquoteright{}d in {FIRE} {III}: The {IMF} in {Quasar} {Accretion} {Disks} from {STARFORGE},} The Open Journal of Astrophysics, 7, \dodoi{10.33232/001c.122857}

\bibitem[{P.~F. Hopkins {et~al.}(2024{\natexlab{b}})Hopkins, Grudic, Su, Wellons, Angles-Alcazar, Steinwandel, Guszejnov, Murray, Faucher-Giguere, Quataert, \& Keres}]{Hopkins2024FORGEda}
Hopkins, P.~F., Grudic, M.~Y., Su, K.-Y., {et~al.} 2024{\natexlab{b}}, \bibinfo{title}{FORGE\textquoteright{}d in {FIRE}: Resolving the {End} of {Star} {Formation} and {Structure} of {AGN} {Accretion} {Disks} from {Cosmological} {Initial} {Conditions},} The Open Journal of Astrophysics, 7, \dodoi{10.21105/astro.2309.13115}

\bibitem[{P.~F. Hopkins {et~al.}(2024{\natexlab{c}})Hopkins, Squire, Su, Steinwandel, Kremer, Shi, Grudic, Wellons, Faucher-Giguere, Angles-Alcazar, Murray, \& Quataert}]{Hopkins2024FORGEdb}
Hopkins, P.~F., Squire, J., Su, K.-Y., {et~al.} 2024{\natexlab{c}}, \bibinfo{title}{FORGE\textquoteright{}d in {FIRE} {II}: The {Formation} of {Magnetically}-{Dominated} {Quasar} {Accretion} {Disks} from {Cosmological} {Initial} {Conditions},} The Open Journal of Astrophysics, 7, \dodoi{10.21105/astro.2310.04506}

\bibitem[{J. Huang {et~al.}(2023)Huang, Lin, \& Shields}]{Huang23}
Huang, J., Lin, D. N.~C., \& Shields, G. 2023, \bibinfo{title}{Metal enrichment due to embedded stars in AGN discs,} Monthly Notices of the Royal Astronomical Society, 525, 5702, \dodoi{10.1093/mnras/stad2642}

\bibitem[{W.-T. Kim \& E.~C. Ostriker(2001)Kim \& Ostriker}]{KimOstriker01}
Kim, W.-T., \& Ostriker, E.~C. 2001, \bibinfo{title}{Amplification, Saturation, and Q Thresholds for Runaway: Growth of Self-Gravitating Structures in Models of Magnetized Galactic Gas Disks,} The Astrophysical Journal, 559, 70, \dodoi{10.1086/322330}

\bibitem[{P.~I. Kolykhalov \& R.~A. Sunyaev(1980)Kolykhalov \& Sunyaev}]{KolykhalovSunyaev1980}
Kolykhalov, P.~I., \& Sunyaev, R.~A. 1980, \bibinfo{title}{The Outer Parts of the Accretion Disks Around Supermassive Black Holes in Galaxy Nuclei and Quasars,} Soviet Astronomy Letters, 357, p. 357

\bibitem[{K. {Kratter} \& G. {Lodato}(2016){Kratter} \& {Lodato}}]{Kratter2016}
{Kratter}, K., \& {Lodato}, G. 2016, \bibinfo{title}{{Gravitational Instabilities in Circumstellar Disks},} \araa, 54, 271, \dodoi{10.1146/annurev-astro-081915-023307}

\bibitem[{Y. {Levin}(2007){Levin}}]{Levin2007}
{Levin}, Y. 2007, \bibinfo{title}{{Starbursts near supermassive black holes: young stars in the Galactic Centre, and gravitational waves in LISA band},} \mnras, 374, 515, \dodoi{10.1111/j.1365-2966.2006.11155.x}

\bibitem[{S. Lizano {et~al.}(2010)Lizano, Galli, Cai, \& Adams}]{Lizano_2010}
Lizano, S., Galli, D., Cai, M.~J., \& Adams, F.~C. 2010, \bibinfo{title}{STABILITY OF MAGNETIZED DISKS AND IMPLICATIONS FOR PLANET FORMATION,} The Astrophysical Journal, 724, 1561, \dodoi{10.1088/0004-637X/724/2/1561}

\bibitem[{L. Lohnert \& A. Peeters(2022)Lohnert \& Peeters}]{LP22}
Lohnert, L., \& Peeters, A. 2022, \bibinfo{title}{Combined dynamo of gravitational and magneto-rotational instability in irradiated accretion discs,} Astronomy \& Astrophysics, 663

\bibitem[{L. Lohnert \& A. Peeters(2023)Lohnert \& Peeters}]{LP23}
Lohnert, L., \& Peeters, A. 2023, \bibinfo{title}{The persistence of magneto-rotational turbulence in gravitationally turbulent accretion disks,} Astronomy \& Astrophysics, 677

\bibitem[{D. Lynden-Bell(1969)Lynden-Bell}]{LyndenBell1969}
Lynden-Bell, D. 1969, \bibinfo{title}{Galactic Nuclei as Collapsed Old Quasars,} Nature, 233, p. 690

\bibitem[{B. {McKernan} {et~al.}(2018){McKernan}, {Ford}, {Bellovary}, {Leigh}, {Haiman}, {Kocsis}, {Lyra}, {Mac Low}, {Metzger}, {O'Dowd}, {Endlich}, \& {Rosen}}]{McKernan2018}
{McKernan}, B., {Ford}, K.~E.~S., {Bellovary}, J., {et~al.} 2018, \bibinfo{title}{{Constraining Stellar-mass Black Hole Mergers in AGN Disks Detectable with LIGO},} \apj, 866, 66, \dodoi{10.3847/1538-4357/aadae5}

\bibitem[{K. Menou \& E. Quataert(2001)Menou \& Quataert}]{MenouQuataert2001}
Menou, K., \& Quataert, E. 2001, \bibinfo{title}{Ionization, Magnetorotational, and Gravitational Instabilities in Thin Accretion Disks Around Supermassive Black Holes,} The Astrophysical Journal, 552, p. 204

\bibitem[{B. Paczy\'nski(1978)Paczy\'nski}]{Paczynski1978}
Paczy\'nski, B. 1978, \bibinfo{title}{A Model of Self-Gravitating Accretion Disk,} Acta Astronomica, 28, p. 91

\bibitem[{V.~I. Pariev {et~al.}(2003)Pariev, Blackman, \& Boldyrev}]{Pariev2003}
Pariev, V.~I., Blackman, E.~G., \& Boldyrev, S.~A. 2003, \bibinfo{title}{Extending the Shakura-Sunyaev approach to a strongly magnetized accretion disc model,} Astronomy and Astrophysics, 407, p. 403

\bibitem[{A. Pawel {et~al.}(1993)Pawel, Lin, \& Wampler}]{Pawel93}
Pawel, A., Lin, D. N.~C., \& Wampler, E.~J. 1993, \bibinfo{title}{Star Trapping and Metallicity Enrichment in Quasars and Active Galactic Nuclei,} Astrophysical Journal, 409, 592

\bibitem[{M.~E. Pessah {et~al.}(2006)Pessah, Chan, \& Psaltis}]{Pessah06}
Pessah, M.~E., Chan, C.-k., \& Psaltis, D. 2006, \bibinfo{title}{The signature of the magnetorotational instability in the Reynolds and Maxwell stress tensors in accretion discs,} Monthly Notices of the Royal Astronomical Society, 372, 183, \dodoi{10.1111/j.1365-2966.2006.10824.x}

\bibitem[{A. Riols \& H. Latter(2018{\natexlab{a}})Riols \& Latter}]{RiolsLatter2018a}
Riols, A., \& Latter, H. 2018{\natexlab{a}}, \bibinfo{title}{Magnetorotational Instability and Dynamo Action in Gravitoturbulent Astrophysical Discs,} MNRAS, 474, p. 2212

\bibitem[{A. Riols \& H. Latter(2018{\natexlab{b}})Riols \& Latter}]{RiolsLatter2018b}
Riols, A., \& Latter, H. 2018{\natexlab{b}}, \bibinfo{title}{Spiral density waves and vertical circulation in protoplanetary discs,} MNRAS, 476, p. 5115

\bibitem[{G. Salvesen {et~al.}(2016)Salvesen, Simon, Armitage, \& Begelman}]{Salvesen2016a}
Salvesen, G., Simon, J.~B., Armitage, P.~J., \& Begelman, M.~C. 2016, \bibinfo{title}{Accretion disc dynamo activity in local simulations spanning weak-to-strong net vertical magnetic flux regimes,} MNRAS, 457, p. 857

\bibitem[{N.~I. Shakura \& R.~A. Sunyaev(1973)Shakura \& Sunyaev}]{ShakuraSunyaev1973}
Shakura, N.~I., \& Sunyaev, R.~A. 1973, \bibinfo{title}{Black Holes in Binary Systems: Observational Appearance,} Astronomy and Astrophysics, 24, p. 337

\bibitem[{J.-M. Shi \& E. Chiang(2014)Shi \& Chiang}]{Shi_2014}
Shi, J.-M., \& Chiang, E. 2014, \bibinfo{title}{GRAVITO-TURBULENT DISKS IN THREE DIMENSIONS: TURBULENT VELOCITIES VERSUS DEPTH,} The Astrophysical Journal, 789, 34, \dodoi{10.1088/0004-637X/789/1/34}

\bibitem[{I. Shlosman \& M. Begelman(1987)Shlosman \& Begelman}]{ShlosmanBegelman1987}
Shlosman, I., \& Begelman, M. 1987, \bibinfo{title}{Self-gravitating accretion disks in active galactic nuclei,} Nature, 329, p. 810

\bibitem[{I. Shlosman \& M. Begelman(1989)Shlosman \& Begelman}]{ShlosmanBegelman1989}
Shlosman, I., \& Begelman, M. 1989, \bibinfo{title}{Evolution of Self-Gravitating Accretion Disks in Active Galactic Nuclei,} The Astrophysical Journal, 341, p. 685

\bibitem[{J. Squire {et~al.}(2025)Squire, Quataert, \& Hopkins}]{Squire2025Rapid}
Squire, J., Quataert, E., \& Hopkins, P.~F. 2025, \bibinfo{title}{Rapid, strongly magnetized accretion in the zero-net-vertical-flux shearing box,} The Open Journal of Astrophysics, 8, \dodoi{10.33232/001c.136467}

\bibitem[{T.~K. Suzuki \& S.-i. Inutsuka(2009)Suzuki \& Inutsuka}]{SuzukiInutsuka2009}
Suzuki, T.~K., \& Inutsuka, S.-i. 2009, \bibinfo{title}{DISK WINDS DRIVEN BY MAGNETOROTATIONAL INSTABILITY AND DISPERSAL OF PROTOPLANETARY DISKS,} The Astrophysical Journal, 691, L49

\bibitem[{D. Syer {et~al.}(1991)Syer, Clarke, \& Rees}]{Syer91}
Syer, D., Clarke, C.~J., \& Rees, M.~J. 1991, \bibinfo{title}{Star–disc interactions near a massive black hole,} Monthly Notices of the Royal Astronomical Society, 250, 505, \dodoi{10.1093/mnras/250.3.505}

\bibitem[{A. Sądowski(2016)Sądowski}]{Sadowski2016}
Sądowski, A. 2016, \bibinfo{title}{Thin accretion discs are stabilized by a strong magnetic field,} Monthly Notices of the Royal Astronomical Society, 459, 4397, \dodoi{10.1093/mnras/stw913}

\bibitem[{T.~A. {Thompson} {et~al.}(2005){Thompson}, {Quataert}, \& {Murray}}]{Thompson2005}
{Thompson}, T.~A., {Quataert}, E., \& {Murray}, N. 2005, \bibinfo{title}{{Radiation Pressure-supported Starburst Disks and Active Galactic Nucleus Fueling},} \apj, 630, 167, \dodoi{10.1086/431923}

\bibitem[{A. Toomre(1964)Toomre}]{Toomre1964}
Toomre, A. 1964, \bibinfo{title}{On the Gravitational Instability of a Disk of Stars,} Astrophysical Journal, 139, p. 1217

\bibitem[{T.~H.~N. Tsung {et~al.}(2025)Tsung, Begelman, Armitage, Jiang, \& Gerling-Dunsmore}]{tsung2025doesmagneticfieldpromote}
Tsung, T. H.~N., Begelman, M.~C., Armitage, P.~J., Jiang, Y.-F., \& Gerling-Dunsmore, H.~J. 2025, \bibinfo{title}{Does magnetic field promote or suppress fragmentation in AGN disks? Results from local shearing box simulations with simple cooling,} \doarXiv{2507.21991}

\end{thebibliography}
\bibliographystyle{aasjournalv7}

\end{document}